\documentclass[letterpaper,twocolumn]{jpsj3}
\usepackage{txfonts}
\usepackage[dvipdfmx]{graphicx,color}
\usepackage{booktabs}
\usepackage{here}
\usepackage{siunitx}

\usepackage{array}
\usepackage{color}
\usepackage{amsmath}
\usepackage{graphicx}
\usepackage{overcite}
\usepackage{txfonts}
\usepackage{bm}

\usepackage[final]{changes}

\definechangesauthor[name = Dai, color=red]{DA}
\definechangesauthor[color=blue]{JF}
\definechangesauthor[color=orange]{FH}
\definechangesauthor[color=green]{HH}

\newcommand{\DA}[1]{\added[id=DA]{#1}}

\newcommand{\UTT}{UTe$_2$}

\newcommand{\Tm}{$T_{\rm m}$}
\newcommand{\Tsc}{$T_{\rm SC}$}
\newcommand{\Hc}{$H_{\rm c2}$}

\newcommand{\Pc}{$P_{\rm c}$}

\newcommand{\Pot}{$P_{\rm O\mbox{-}T}$}

\title{
Pressure-induced structural phase transition and new superconducting phase in \UTT
}

\author{
Fuminori~Honda$^{1,2}$\thanks{honda.fuminori.790@m.kyushu-u.ac.jp},
Shintaro~Kobayashi$^{3}$,
Naomi~Kawamura$^{3}$,
Saori I.~Kawaguchi$^{3}$,
Takatsugu~Koizumi$^{2}$,
Yoshiki~J.~Sato$^{2,4}$,
Yoshiya~Homma$^{2}$,
Naoki~Ishimatsu$^{5}$,
Jun~Gouchi$^{6}$,
Yoshiya~Uwatoko$^{6}$,
Hisatomo~Harima$^7$,
Jacques~Flouquet$^8$,
and Dai~Aoki$^2$
}

\inst{
$^{1}$Central Institute of Radioisotope Science and Safety, Kyushu University, Fukuoka 819-0395, Japan\\
$^{2}$Institute for Materials Research, Tohoku University, Oarai, Ibaraki 311-1313, Japan\\
$^{3}$Japan Synchrotron Radiation Research Institute, Sayo, Hyogo 679-5198, Japan\\
$^{4}$Department of Physics, Faculty of Science and Technology, Tokyo University of Science, Noda, Chiba 278-8510, Japan\\
$^{5}$Graduate School of Advanced Science and Engineering, Hiroshima University, Higashihiroshima, Hiroshima 739-8526, Japan\\
$^{6}$Institute for Solid State Physics, University of Tokyo, Kashiwa, Chiba 277-8581, Japan\\
$^7$Graduate School of Science, Kobe University, Kobe 657-8501, Japan\\
$^8$Universit\'{e} Grenoble Alpes, CEA, Grenoble INP, IRIG, PHELIQS, F-38000 Grenoble, France
}

\abst{
We report on the crystal structure and electronic properties of the heavy fermion superconductor UTe$_2$ at high pressure up to 11 GPa, as investigated by X-ray diffraction and electrical resistivity experiments.
The X-ray diffraction measurements under high pressure using a synchrotron light source reveal anisotropic linear compressibility of the unit cell up to 3.5 GPa, while a pressure-induced structural phase transition is observed above $P_{\rm O\mbox{-}T}\sim 3.5-4\, {\rm GPa}$ at room temperature, where the body-centered orthorhombic crystal structure with the space group $Immm$ changes into a body-centered tetragonal structure with the space group $I4/mmm$.
The molar volume drops abruptly at $P_{\rm O\mbox{-}T}$, while the distance between the first-nearest neighbor of U atoms, $d_{\rm U\mbox{-}U}$, increases, implying a switch from the heavy electronic states to the weakly correlated electronic states.
Surprisingly, a new superconducting phase at pressures higher than 7 GPa was detected at $T_{\rm sc}> 2\,{\rm K}$ with a relatively low upper-critical field, $H_{\rm c2}(0)$. 
The resistivity above $3.5\,{\rm GPa}$, thus, in the high-pressure tetragonal phase, shows a large drop below $230\,{\rm K}$, which may also be related to a considerable change from the heavy electronic states to the weakly correlated electronic states.
}

\begin{document}

\maketitle

\section{Introduction}

\UTT\ is one of the hottest topics in condensed matter physics, because of its unusual superconducting properties, such as topological superconductivity, field-reentrant superconductivity, and multiple superconducting phases in a magnetic field as well as under high-pressure \cite{Aoki2019, Ran2019, Daniel2019, Aoki2020JPSJa, Knebel2019,Ran2019_NaturePhys,Ran2020PRB}. 
A possible analogy to ferromagnetic superconductors, such as UGe$_2$,\cite{Saxena2000} URhGe,\cite{Aoki2001} and UCoGe\cite{Huy2007}, had been pointed out at the beginning of the discovery of superconductivity in \UTT. 

Indeed, the markedly high field-reentrant superconductivity detected in the hard-magnetization axis in UCoGe \cite{Huy2007} and URhGe \cite{Aoki2001} resembles that observed in \UTT\ for the $H \parallel b$-axis. 
On the other hand, no magnetic ordering was found at low temperatures, revealing a paramagnetic ground state in \UTT,\cite{Sundar2019} which is quite in contrast to the above-mentioned ferromagnetic superconductors. 

In UTe$_2$, a first-order metamagnetic transition is observed at $H_{\rm m}\sim 35\,{\rm T}$ at low temperatures~\cite{Miyake2019, KnafoJPSJ2019, Ran2019, Knebel2020, Miyake2021}. 
Although \UTT\ had been considered on the verge of ferromagnetic ordering, 
no microscopic evidence for ferromagnetic interactions was obtained experimentally.
On the other hand, inelastic neutron scattering experiments clearly showed the antiferromagnetic fluctuations at the incommensurate wave vector,~\cite{Duan2020, Knafo2021}.
Recently, STM experiments have led to the observation of the charge density wave (CDW)~\cite{Aishwarya2022} or the pair density wave (PDW)~\cite{Gu2022} in the superconducting state.
After three years of studies, UTe$_2$ is found to be a remarkable heavy-fermion system, where the electronic, magnetic, and charge instabilities coupled to the occupancy of the 5$f$ levels play important roles.~\cite{Aoki2022_review}

%%%%%%%%%%%%%%%   fig01  %%%%%%%%%%%%%%%%%%%%%%%%
\begin{fullfigure}[hbt]
\begin{center}
\includegraphics[width=1\hsize]{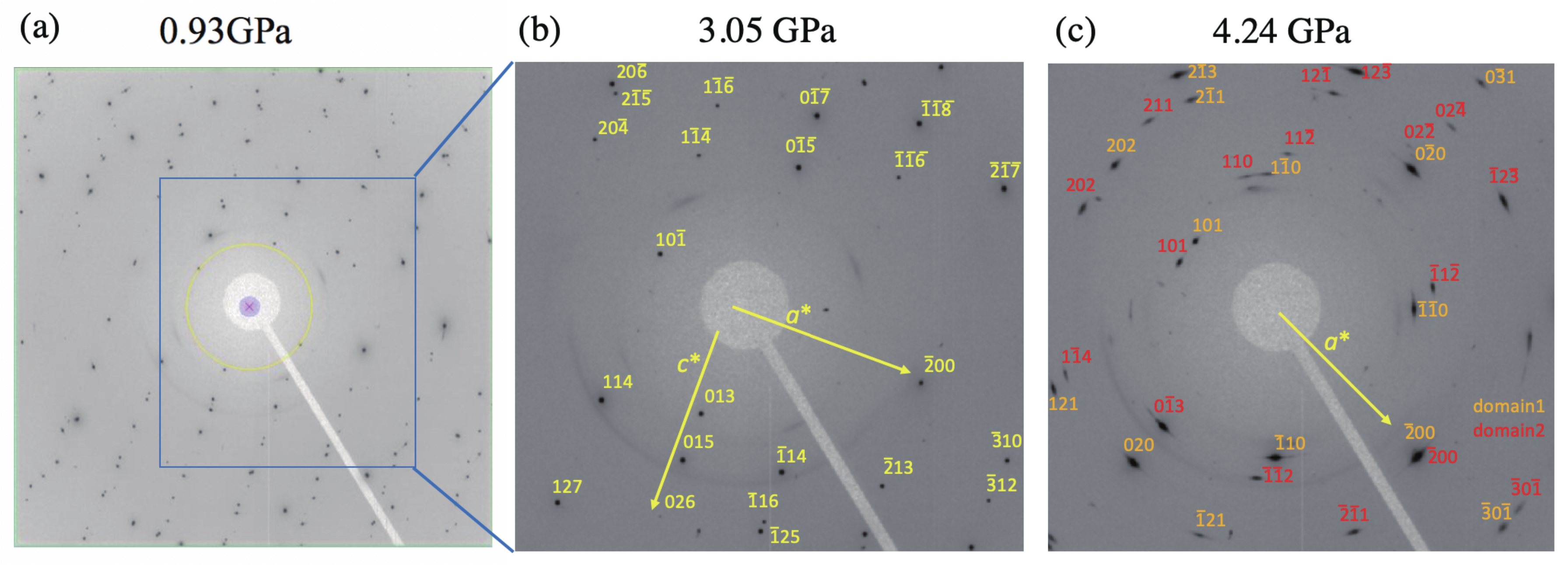}
\caption{(Color online) 
Single-crystal XRD patterns at (a) 0.93 GPa, (b) 3.05 GPa with the low-pressure orthorhombic phase, and (c) 4.24 GPa with the high-pressure tetragonal phase measured at $300\,{\rm K}$. The corresponding Miller indices are indicated in panels (b) and (c) on the basis of the orthorhombic and tetragonal structures, respectively. The crystal orientations in the reciprocal space are also shown by arrows. In panel (c), XRD patterns are indexed on the basis of two different domains, ``domain 1'' and ``domain 2''.
\label{Laue}}
\end{center}
\end{fullfigure}
%%%%%%%%%%%%%%%%%%%%%%%%%%%%%%%%%%%%%%%

To date, the relationship between the crystal structure and electronic properties is intensively discussed. 
Hence, we recall the crystal structure of \UTT. 
\UTT\ crystallizes in the body-centered orthorhombic structure with the space group $Immm$ (No. 71, $D_{25}^{2h}$)~\cite{Ikeda2006}, where four formula units are included in the ``conventional unit cell'' (Z = 4). 
%Hereafter we just call ``unit cell''.}
The nearest and next nearest neighbor U atoms locate along the $c$- and $a$-axes forming a rectangular shape within the $ac$-plane. 
It is also pointed out that the nearest U forms a two-leg ladder along the $a$-axis. 
The crystal structure is of the so-called UTe$_2$ type, and few other materials form this structure, 
demonstrating that it is a quite unique structure.
Note that further lowering the symmetry from the space group $Immm$ in the electronic system has been theoretically proposed~\cite{Harima2022}, 
as it is known in the ``hidden-order'' compound URu$_2$Si$_2$.

The superconducting properties of \UTT\ depend strongly on its sample quality \cite{AokiJPSConfSer2020, Cairns2020,Aoki2022_review}. 
It is also claimed that a significant deficiency of U rather than Te affects the electronic/superconducting properties of \UTT, as shown by their precise analysis, indicating that there may be an instability in the crystal structure of \UTT.\cite{Haga2022}

The pressure response of \UTT\ is very sensitive, displaying different superconducting and magnetic states. 
After applying pressure, the superconducting transition temperature splits above 0.3 GPa, revealing multiple superconducting phases. At around 1 GPa, an unusual enhancement of \Hc\ for the $H\parallel a$-axis was detected at the low-temperature and high magnetic field region as a consequence of the multiple superconducting phases \cite{AokiJPSJ2021, Li2021}.
Even at ambient pressure, multiple superconducting phases were also detected for the $b$-axis in specific heat measurements associated with the field-reentrant superconducting state. 
Above the critical pressure \Pc\ of 1.5 GPa, superconductivity is suppressed, and the magnetically ordered phase, most likely antiferromagnetism, appears under higher pressure \cite{Daniel2019, Li2021}. 
%Moreover, it is also reported that the non-superconducting \UTT\ sample at ambient pressure, which is most likely due to the poor sample quality as indicated by the large residual resistivity at low temperatures, shows pressure-induced superconductivity coexisting with the magnetically ordered state under quasi-uniaxial pressure condition \cite{Yang2022}.

It is surprising that \UTT\ reveals such rich and unexpected phase diagrams under pressure and a magnetic field. 
This is most likely due to a singularity of the crystal structure and instability of electronic states. 
It is important to elucidate the pressure response of the crystal structure and the electronic states at high pressures with good hydrostatic conditions using a high-quality single crystal. 
Thus, we performed X-ray diffraction (XRD) experiment up to 11 GPa and electrical resistivity measurement up to 9 GPa.
We found a considerable change in the crystal structure from the orthorhombic structure (space group: $Immm$) to the tetragonal structure ($I4/mmm$) above the critical pressure $P_{\rm O\mbox{-}T}\sim 3.5-4\, {\rm GPa}$ at room temperature.
Furthermore, the resistivity above 3.5 GPa shows a new phase transition at around 230 K, which is almost unchanged with increasing pressure up to 9 GPa. 
In addition, a reappearance of superconductivity was found at 9 GPa with low \Hc, indicating weak electronic correlations, in marked contrast to superconductivity at low pressures. 
\DA{These results have been briefly reported in a JPS meeting and LT29 by oral and poster presentations, respectively.~\cite{Honda_meeting} 
}

\section{Experimental Procedure}

Single crystals of \UTT\ were grown by the chemical vapor transport method using iodine as a transport agent as described in ref.~\citen{Aoki2019}. 
The quality of the single crystals was checked on the basis of the sharpness of their X-ray Laue patterns.
The residual resistivity ratios at ambient pressure were $20$--$30$, and a sharp specific heat jump due to the superconducting transition was detected at $1.6\,{\rm K}$.

The XRD experiments under high pressure were carried out on the BL10XU beamline at JASRI/SPring-8, Japan, using a wavelength of $\lambda$ = 0.4130 \AA. 
%Since UTe$_2$ is easily oxidized or damaged, particularly for minute pieces of samples, special attention was paid.
Particular care was needed for the minute pieces of samples as UTe$_2$ is easily oxidized and damaged.
Three XRD experiments using diamond anvil cells (DACs) were performed, where two of them were carried out on single crystals with different orientations, and the other was done using a powder made from pulverized single crystals from the same batch.
%the other was performed using a powder of a pulverized crystal from the same batch of the single crystals.
For the experiments using single crystals, two single crystals of \UTT, namely, a crushed single crystal piece, having a (110) plane with a size of 90 $\mu$m $\times$ 90 $\mu$m and a thickness of 30 $\mu$m (labeled as ``single (1st)''), and an oriented crystal having a (100) plane with a size of 120 $\mu$m $\times$ 80 $\mu$m and a thickness of 23 $\mu$m by polishing (labeled as ``single (2nd)'') were loaded in different DAC setups with SUS gaskets together with helium as the pressure-transmitting medium to obtain compensated diffraction data. XRD measurement of a powder sample was also performed using helium as a pressure-transmitting medium. The pressure inside the DACs was measured using the conventional ruby scale.
The pressure values shown are the average of the pressure before and after the diffraction measurements; the error is based on the difference between them.
The sample pressure was regulated by a helium gas compression system. An imaging plate was used to obtain two-dimensional (2D) diffraction patterns via oscillation photography methods, and the resulting data were integrated along the radial direction to obtain a one-dimensional (1D) diffraction pattern.

Electrical resistivity under high pressures up to 9 GPa was measured by a conventional four-probe method \DA{with the current along the $a$-axis}, using a Palm Cubic anvil press system \cite{Cheng2014} installed at the Institute of Solid State Physics, The University of Tokyo in a $^4$He cryostat down to $2\,{\rm K}$.
A mixture of Fluorinert FC70 and FC 77 at a 1:1 ratio was used as a pressure-transmitting medium. Good hydrostaticity of the pressure condition in this cubic anvil press was mainly guaranteed by the system itself, where the Teflon capsule inside the cubic pyrophyllite and/or MgO gasket was compressed from six directions simultaneously.
\DA{At $9\,{\rm GPa}$, the lowest temperature was further extended down to $30\,{\rm mK}$ in a dilution refrigerator, and the magnetic field was applied along the $c$-axis of the original orthorhombic structure.
}

%%%%%%%%%%%%%%%   fig02  %%%%%%%%%%%%%%%%%%%%%%%%
\begin{fullfigure}[tbh]
\begin{center}
\includegraphics[width=0.6\hsize]{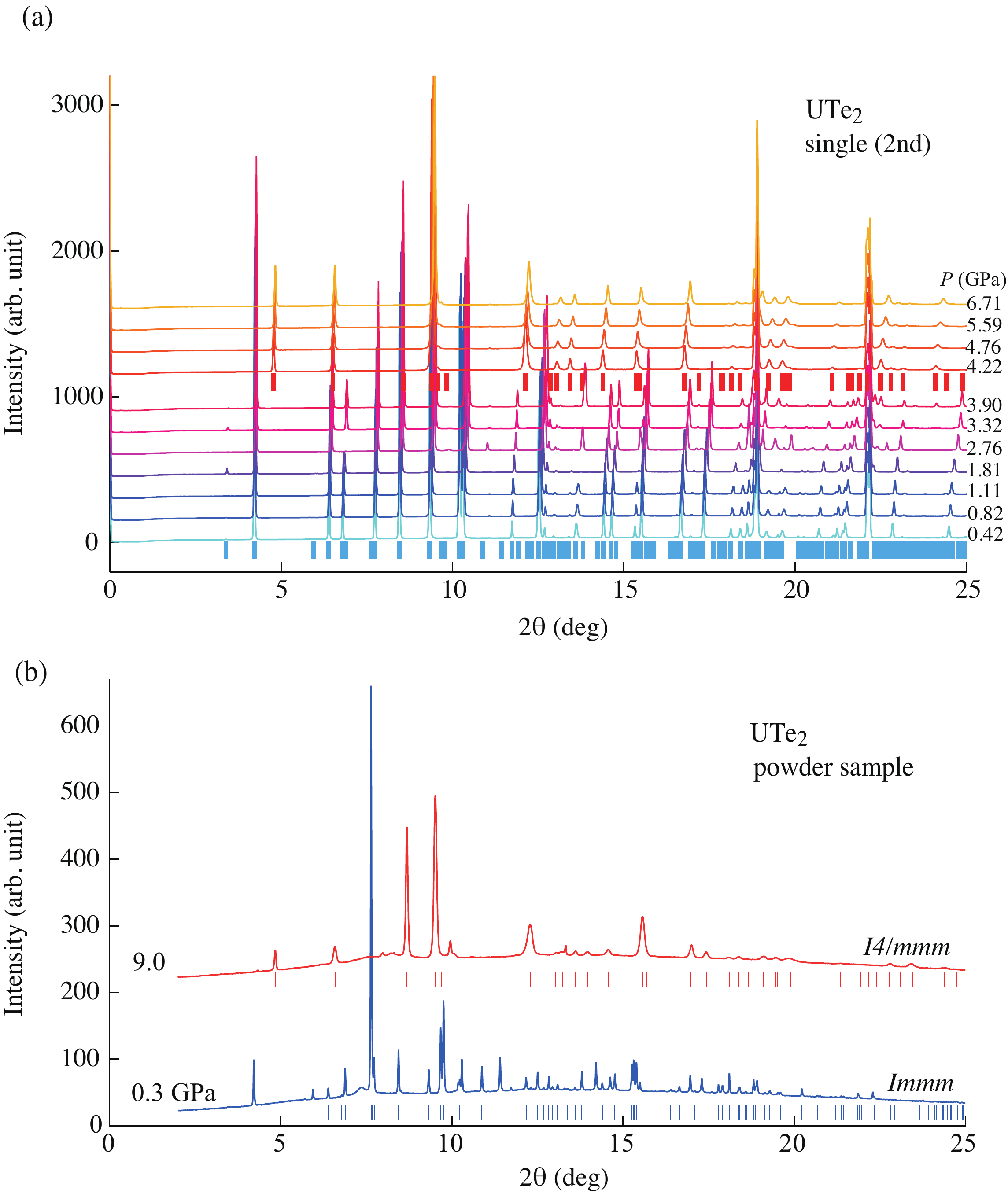}
\caption{(Color online) 
(a) XRD patterns of a \UTT\ single crystal ``single (2nd)'' at 300 K under high pressure. The blue and red bars below the data of 0.42 and 4.22 GPa indicate peak positions calculated according to the obtained lattice parameters for the low- and high-pressure phases, respectively. (b) XRD patterns of powder samples at 0.3 and 9.0 GPa, where vertical bars in blue and red are the peak positions calculated on the basis of orthorhombic ($Immm$) and tetragonal ($I4/mmm$) structures, respectively.
\label{powXRD}}
\end{center}
\end{fullfigure}
%%%%%%%%%%%%%%%%%%%%%%%%%%%%%%%%%%%%%%%

%%%%%%%%%%%%%%%   fig03  %%%%%%%%%%%%%%%%%%%%%%%%
\begin{fullfigure}[t]
\begin{center}
\includegraphics[width=0.6\hsize]{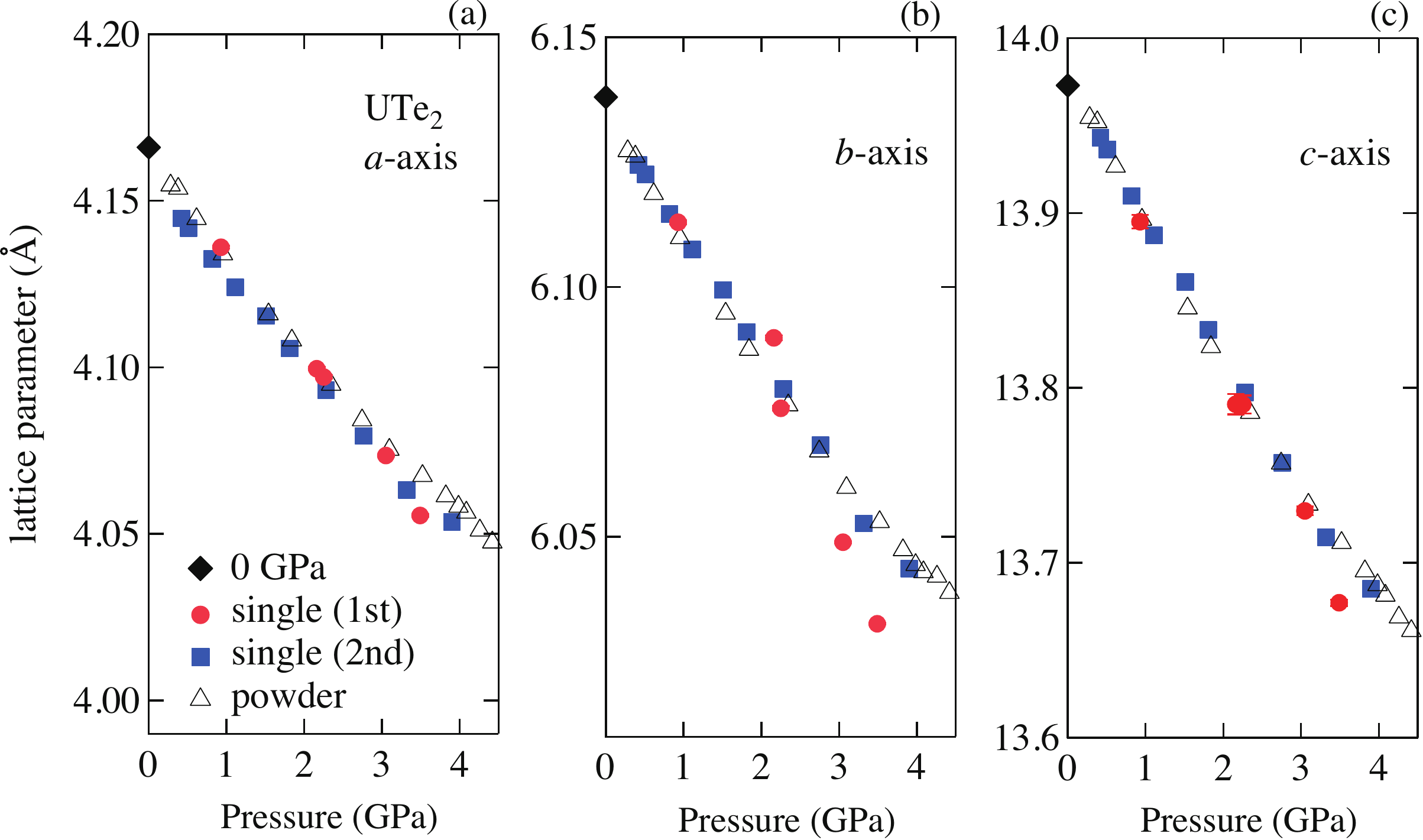}
\caption{(Color online)
Pressure dependence of lattice parameters for orthorhombic structures at low pressures. 
Open and closed symbols indicate the results obtained from single crystals and powder samples, respectively. 
Data at ambient pressure are taken from Ref.~\citen{Ikeda2006}. 
\label{abcDEP}}
\end{center}
\end{fullfigure}
%%%%%%%%%%%%%%%%%%%%%%%%%%%%%%%%%%%%%%%

\section{Experimental Results and Analyses}

\subsection{X-ray diffraction under high pressure}

Figure~\ref{Laue} shows the XRD patterns of a crushed \UTT\ single crystal (``single (1st)'') at (a) 0.93, (b) 3.05, and (c) 4.24 GPa at room temperature. To obtain more diffraction spots, the pressure cell was swung by $\pm$ 10 deg around the origin. Here, Figs.~\ref{Laue}(b) and \ref{Laue}(c) are focused on the magnified rectangular area in Fig.~\ref{Laue}(a) for clarity.
The diffraction patterns at 0.93 and 3.05 GPa are well reproduced by the diffraction patterns of the reported orthorhombic structure with the space group $Immm$.
The Miller indices of each diffraction spot are indicated on the basis of the reported structure in Fig.~\ref{Laue}(b). Directions of the approximate reciprocal vectors are also indicated by arrows. 
On the other hand, it is easily recognized that the diffraction pattern at 4.24 GPa has changed considerably. 
Note that the diffraction pattern changed during the measurement at 3.49 GPa; thus, the structural phase transition occurs at this pressure on the ``single (1st)'' sample.
Upon compression of the sample above 3.5 GPa, the number of Bragg peaks is almost doubled. 
This indicates that the sample has transformed into a multi-domain and/or the lattice symmetry has decreased. 
The observed Laue spots are clearly broadened and start to form the so-called Debye$-$Scherrer rings (i.e., their shapes become ellipsoidal). 
The Miller indices of each diffraction spot in Fig.~\ref{Laue}(c) are indicated on the basis of the tetragonal structure with the space group $I4/mmm$ including two domains; the analysis of the structural phase transition is explained in more detail later.

The obtained 2D diffraction patterns are converted into 1D intensity $2 \theta$ angle profiles by integrating along the radial direction from the beam center in order to analyze unit cell parameters. 
Figure~\ref{powXRD}(a) shows the diffraction peaks of a \UTT\ single crystal with the (110) plane (``single (2nd)'') under high pressures at room temperature.
Since the patterns are obtained from a single crystal (single domain), not all the diffracted peaks expected from the orthorhombic crystal structure with the space group $Immm$ were observed, and the relative heights of the peak intensities were not well reproduced, but they were sufficient to determine lattice parameters. 
It is again clear that the diffraction patterns have changed at pressures above 4 GPa. 
Above 4 GPa, the observed XRD pattern was indexed to a body-centered tetragonal system to give the following crystal structure parameters: $a$ = 3.89 \AA\ and $c$ = 9.80 \AA. 
The detailed crystal structure analysis of the high-pressure tetragonal phase is described in the following section. 
The $a$-axis and $c$-axis directions in the low-pressure phase turn to a nearly $a$ (and $b$)-axis direction in the high-pressure tetragonal phase through the structural phase transition, which means that the original $a$-axis in the orthorhombic structure almost maintains its orientation in the high-pressure tetragonal phase.
To confirm the structural change from $Immm$ to $I4/mmm$, the obtained XRD patterns using powder samples at 0.3 and 9.0 GPa were displayed in Fig.~\ref{powXRD}(b).
%The experimental results at 0.3 and 9GPa were well explained by the calculated peak positions based on the obtained lattice parameters with the low-pressure orthorhombic ($Immm$) and high-pressure tetragonal ($I4/mmm$) phases, respectively.
The diffraction patterns at 0.3 and 9.0 GPa were well explained on the basis of the calculated peak positions with the lattice parameters of the low-pressure orthorhombic ($Immm$) phase with $a$ = 4.155, $b$ = 6.127, and $c$ = 13.954 \AA\ and the high-pressure tetragonal ($I4/mmm$) phase with $a$ = 3.859 and $c$ = 9.775 \AA, respectively. 
%\DA{The pressure-induced tetragonal structure is retained, at least, up to our highest pressure, 11 GPa.}
The pressure-induced tetragonal structure was found to be maintained at least up to 11 GPa, the highest pressure measured in this study.

From these results, we determined the pressure dependence of the lattice parameters of \UTT\ from both single crystals and powder samples.  
The analyzed unit cell parameters of the orthorhombic $a$-, $b$-, and $c$-axes as functions of pressure are shown in Figs.~\ref{abcDEP}(a) $-$ ~\ref{abcDEP}(c), respectively.
%Figs.~\ref{abcDEP}(a), ~\ref{abcDEP}(b), and ~\ref{abcDEP}(c), respectively. 
%The data at ambient pressure shown by open-diamond are taken from Ref.\citen{Ikeda2006}. 
%For the pressure data, parameters obtained on single crystal and powder samples are shown in closed and open symbols, respectively. 
The results from single crystals and powder samples are basically the same.
The lattice parameters along the $a$-, $b$-, and $c$-axes decrease almost linearly with increasing pressure. 
The compression curves are anisotropic, and the obtained linear compressibility ($k_i =-(1/L_{i0}) (dL_i / dp)_{p=0}$, $i$ = $a$, $b$, $c$-axes), where $L_i$ is a lattice parameter along the crystallographic $i$-axis, are $k_a$ = 7.2 $\times$ 10$^{-3}$, $k_b$ = 4.0 $\times$ 10$^{-3}$, and $k_c$ = 5.8 $\times$ 10$^{-3}$ (GPa$^{-1}$) for $a$-, $b$-, and $c$-axes, respectively. The $a$-axis is the most compressive and the $b$-axis is the hardest. 
The elastic properties of uranium compounds have not been well established so far, owing to a lack of sufficient examples. However, it is suggested empirically that the direction of U$-$U bonds is distinctly the ``soft'' crystallographic direction in the UTX (T, transition metal; X, metalloid) system, since the 5$f$-electrons of U participate in the bonding.\cite{Havela2001, Maskova2012} In \UTT, the directions along the $c$- and $a$-axes correspond to the 1st and 2nd nearest neighbor U-bondings, respectively, and show a higher linear compressibility than that along the $b$-axis, which is consistent with the above-mentioned hypothesis.
The calculated volume compressibility $k_V$ = $k_a$ + $k_b$ + $k_c$ = 17 $\times$ 10$^{-3}$ (GPa$^{-1}$) corresponds to the bulk modulus of $B$ = 59 GPa. 
This value is even smaller than that of the tellurium element itself, $B$ = 65 GPa.
\UTT\ is very ``soft''. 
Thus, considerable changes in the physical properties of \UTT\ under pressure are caused by such a small bulk modulus.

\begin{figure}[htb]
\begin{center}
\includegraphics[width=\hsize]{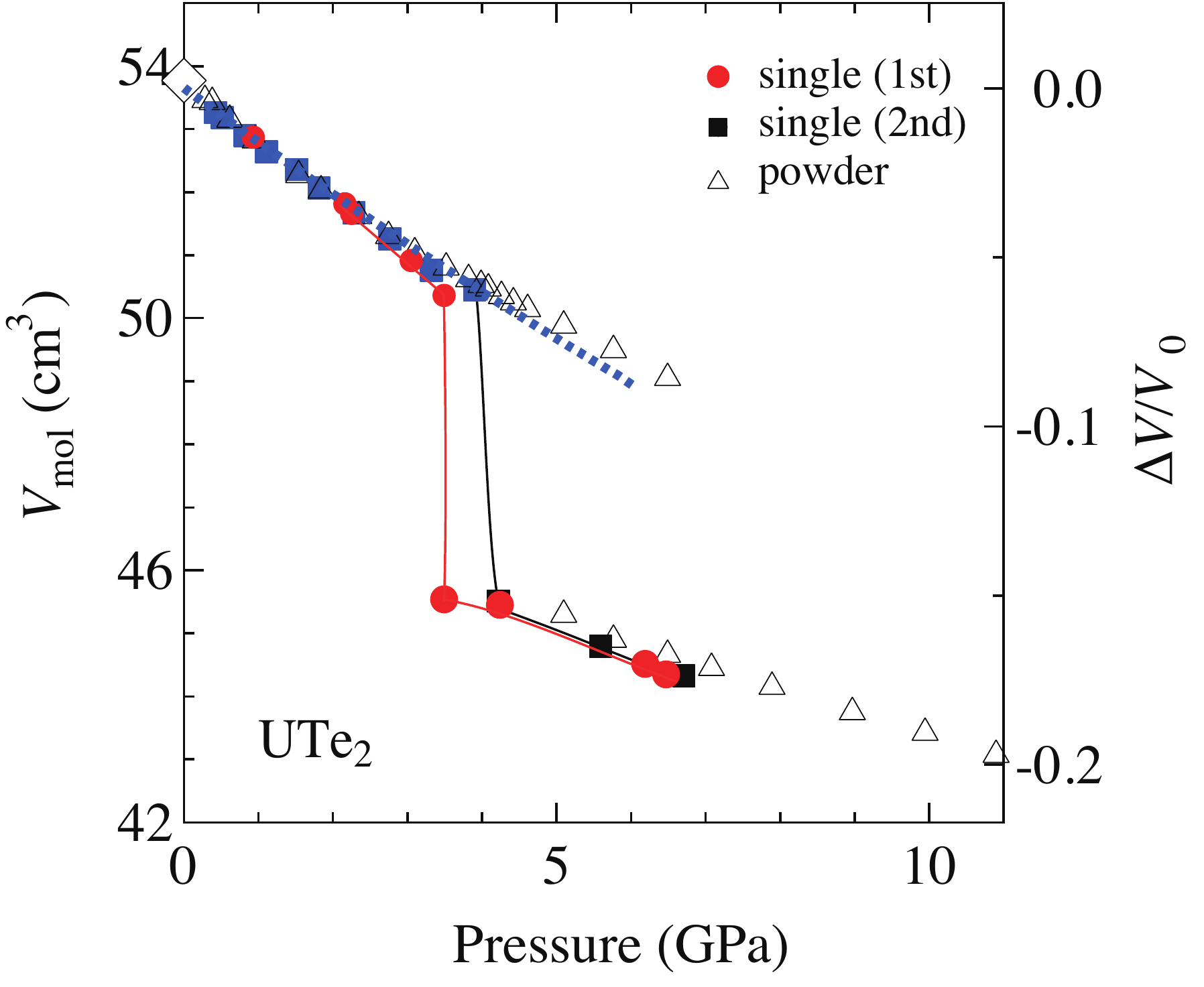}
\end{center} 
\caption{(Color online)
Pressure dependence of the molar volume in UTe$_2$ based on the orthorhombic ($Immm$) and tetragonal ($I4/mmm$) structure at $300\,{\rm K}$. Closed and open symbols are the results from single crystals and powder samples, respectively. An open diamond is taken from Ref.~\citen{Ikeda2006}.
The right axis corresponds to the scaled volume change. 
The dotted blue line indicates the results of fitting by the Murnaghan equation of state.
The solid red and black lines are guides to the eye.}
\label{AVdep}
\end{figure}

Figure~\ref{AVdep} shows the molar volume change as a function of pressure up to 11 GPa at room temperature in single crystals (closed symbol) and powder samples (open symbol). The pressure dependence of the relative volume change ($\Delta V/V_0 = (V(P)-V_0)/V_0)$ with respect to the volume at ambient pressure ($V_0$) reaches 3.5\% at 2 GPa, which is comparable to the volume reduction in liquid $^{3}$He (about 5\%), where the paramagnetic Fermi liquid phase changes to the solid phase on the melting curve at a low temperature. 
Thus, one can easily imagine the rather considerable change that occurs in the electronic properties of \UTT\ such as the disappearance of its superconductivity and the occurrence of magnetic ordering at about 2 GPa.

Volume compression is analyzed using the following Murnaghan equation of state, 
\begin{equation}
\frac{V}{V_0} =   \left [   P   \left(  \frac{B_0^{'}}{B_0} \right) + 1     \right]  ^{- 1/B_0^\prime},
%= \{ P \frac{B_0^'}{B_0} +1  \} ^{- \frac{1}{B_0^'}},
\label{BM}
\end{equation}
where $B_0$ and $B_0^{'}$ are the bulk modulus and its pressure derivative at ambient pressure, respectively, and $V_0$ was calculated from the extrapolated lattice parameters to 0 GPa from the present pressure data. The best-fit parameters are $B_0$ = 59 GPa and $B_0^{'}$ = 1.9.
The dotted blue line in Fig.~\ref{AVdep} is the calculated curve using these parameters and Eq. (\ref{BM}). 
The molar volume drops at the structural phase transition with a volume change of 10\%.
Note that the crystal structure returned to the original orthorhombic structure below 1 GPa during decreasing pressure, which indicates that the latent heat at the structural transition is large.

The results of single crystals (closed symbol) and powder samples (open symbol) are basically similar, but the noted differences are the critical pressure of structural phase transition and coexistence region of low- and high-pressure phases. 
The single crystal data show the sharp transition to the tetragonal structure at $P_{\rm O\mbox{-}T} = 3.5$ and $4 \,{\rm GPa}$, while the coexistence of two structures occurs between 5 and $7\,{\rm GPa}$ in the powder sample.
This is most likely due to the pressure distribution in the gasket and internal strain and/or defects that could could happen in the pulverization of single crystals.
%%%%%%%%%%%%%%%   fig04  %%%%%%%%%%%%%%%%%%%%%%%%
%%%%%%%%%%%%%%%%%%%%%%%%%%%%%%%%%%%%%%%%

\subsection{Electrical resistivity under high pressure}

\begin{figure}[htb]
\begin{center}
\includegraphics[width=0.8\hsize]{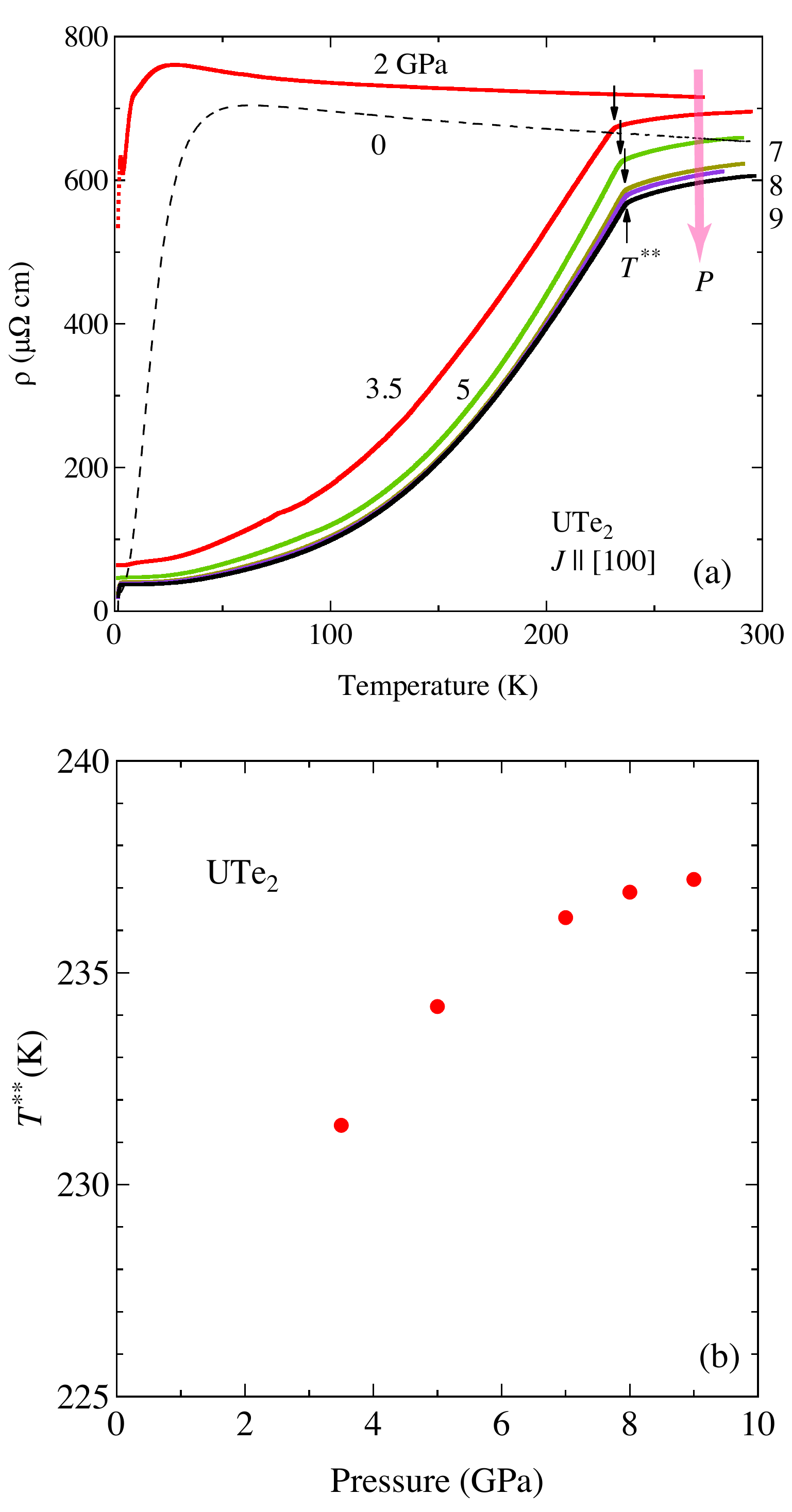}
\end{center} 
\caption{(Color online) 
(a) Electrical resistivity at different pressures up to $9\,{\rm GPa}$ for $J \parallel a$-axis in UTe$_2$. The data at ambient pressure shown by the dashed line are taken from Ref.~\citen{Aoki2019}.
(b) Pressure dependence of $T^{\ast\ast}$.
}
\label{HPRT}
\end{figure}

Figure~\ref{HPRT}(a) shows the temperature dependence of the electrical resistivity $\rho$($T$) of the \UTT\ single crystal with the current along the orthorhombic $a$-axis \DA{at different pressures. 
The $\rho$($T$) curve in \UTT\ at ambient pressure is characterized by almost temperature independence above 100 K and an abrupt decrease in $\rho$ below 50 K, as shown in the dashed line in Fig.~\ref{HPRT}(a). 
Note that the bare LDA band structure calculation predicts a Kondo semiconductor despite the fact that the relatively large volumes of Fermi surfaces (about 20\% in BZ for each hole and electron Fermi surface with the compensation) are experimentally detected.~\cite{Aoki2022} 
The Coulomb repulsion $U$ and the strong correlation must be taken into account in the calculations~\cite{Ishizuka2019} to explain real Fermi surfaces, suggesting that the electronic states are sensitive to pressure and temperature. 
At 2 GPa, $\rho$($T$) still resembles the curve at ambient pressure with a gradually enhanced hump around 25 K, indicating that the heavy electronic state is still preserved or even enhanced at low temperatures. 
At lower temperatures, two anomalies are detected at 8.5 and 2.8 K, which may correspond to the transitions for the pressure-induced WMO and MO phases, respectively~\cite{Li2021,AokiJPSJ2021}.
The MO and WMO phases are suggested to be ``magnetic ordered'' and ``weakly magnetic ordered'' phases, respectively. The former is most likely a pressure-induced magnetically ordered state, and the origin of the latter state is believed ${not}$ driven by the long-range magnetic order but by some short-range correlations.
At 3.5 GPa, the $\rho$($T$) behavior changes considerably, where the electrical resistivity decreases monotonically with decreasing temperature and shows a clear shoulder at around $T^{\ast\ast} = 230 {\rm K}$ and a small upturn at around \Tm\ = 4 K.
The anomaly at \Tm\ is consistent with the pressure-induced magnetic transition reported previously \cite{Daniel2019, Aoki2020JPSJa,Li2021}.
Note that $\rho$($T$) at 3.5 GPa was measured after driving the pressure up to 3.5 GPa at room temperature, indicating that the system has already transformed to the high-pressure tetragonal phase. Because of the huge hysteresis, once the system enters the high-pressure phase, it maintains its structure down to the lowest temperature.
With increasing pressure above 5 GPa, the \Tm\ anomaly smeared out, and $T^{\ast\ast}$ slightly increases monotonically with an initial slope of 1.9 K/GPa as shown in Fig.~\ref{HPRT}(b). 
The origin of the anomaly at $T^{\ast\ast}$ has been unclear so far, but one can speculate that it corresponds to a considerable change from a single site behavior to a well-coupled array at relatively high temperatures,
which is basically one order of magnitude higher than the crossover temperature $T^\ast$ detected at low pressure.}
Note that the resistivity at room temperature as a function of pressure shows no drastic change at the critical pressure $P_{\rm O\mbox{-}T}$ for structural transition. 

Prior to the present measurement, we carried out electrical resistivity measurements under high pressure twice. Each time, we have found a sudden decrease in electrical resistivity at low temperatures above 6 GPa.
Figure~\ref{HPSC}(a) shows the low-temperature part of $\rho$($T$) at pressures higher than 3.5 GPa in the high-pressure tetragonal phase. 
The sudden decrease in $\rho$($T$) is again confirmed above 7 GPa, which is indicative of superconductivity as described later.
The onset of the pressure-induced superconductivity is detected at 7 GPa below 2 K. At 9 GPa, a clear superconducting transition by zero resistivity was detected below about 1 K.
%, using a Palm Cubic anvil cell in a dilution fridge. 
To investigate the low-temperature properties in more detail, resistivity experiments were extended at very low temperatures down to $30\,{\rm mK}$.
The onset of the superconducting transition temperature \Tsc\ is 2.2 K at 9 GPa in a zero magnetic field, where the superconducting transition temperature is defined with 80 \% of the residual resistivity. 
Note that the broadened SC transition can be attributed to the strain and cracks arising due to the structural transition associated with the drastic change in lattice parameters. 
In fact, the sample unloaded from the Palm Cubic anvil pressure cell after the measurement breaks up into small pieces, which was also seen in the samples taken after the XRD measurements under pressure. 

We further investigated the superconducting property at 9 GPa in a magnetic field and constructed the phase diagram. 
Here, the magnetic field was applied along the original $c$-axis in the low-pressure orthorhombic phase. %, which corresponds to the direction tilted by about 25 deg from $a$ to $c$-axis in the high-pressure tetragonal phase. 
Figures~\ref{HPSC}(b) and ~\ref{HPSC}(c) show the temperature dependence of electrical resistivity at the different magnetic fields and magnetic field dependence at 30 mK, respectively. 
Pressure-induced superconductivity disappears above 2.5 T, although the onset of small superconductivity persists up to about 3.8 T. We show in Fig.~\ref{HPSC}(d) the temperature dependence of the upper critical field \Hc\ defined as 80 \% of the resistivity of the normal state. 
The slope of the upper critical field is $-dH_{c2}/dT \sim 1\, {\rm T/K}$ at \Tsc\ = 2.2 K, and \Hc ($T$=0) is about 2.5 T, which is smaller than the Pauli limit based on the weak coupling BCS theory. 
Furthermore, the temperature dependence of electrical resistivity follows the Fermi liquid behavior ($\rho = \rho_0+AT^2$) with a small $A$ coefficient ($A_{9{\rm GPa}}\sim 1.4 \times 10^{-3}\,\mu\Omega\cdot {\rm cm/K^2}$), indicating that superconductivity is based on the weakly correlated electronic states without any contribution from 5$f$-electrons.
This can be compared with the $A$ coefficient at ambient pressure, $A_0\sim 0.9\,\mu\Omega\!\cdot\!{\rm cm/K^2}$.~\cite{Aoki2019}
The ratio $A_{\rm 9GPa}/A_0\sim 600$ is of the same order as $(T^{\ast\ast}/T^\ast)^2 \sim 300$.
Furthermore, the Sommerfeld coefficient for specific heat is expected to be less than $10\,{\rm mJ\, K^{-2}mol^{-1}}$ at high pressures,
assuming the Kadowaki$-$Woods ratio, suggesting a weak electronic correlation.
The small initial slope of $H_{\rm c2}$ at $9\,{\rm GPa}$ is also reasonable, compared with the large values, $34\,{\rm T/K}$ and $7.5\,{\rm T/K}$ for the $b$- and $c$-axes, respectively, detected at ambient pressure,
in accordance with the change in the effective masses,
since $H_{\rm c2}$ is proportional to $(m^\ast T_{\rm sc})^2$ assuming that the Fermi surfaces are unchanged.

The conventional superconductivity of the tellurium element is induced above 4 GPa, but the superconducting state of tellurium is easily eliminated at the magnetic field of 0.03 T. Thus, the observed pressure-induced superconductivity is not due to the remaining tellurium but is intrinsic in the high-pressure tetragonal phase of \UTT.
To clarify that the superconducting state originates from the bulk, ac susceptibility and specific heat measurements under high pressure are in progress.

\begin{figure}[htb]
\begin{center}
\includegraphics[width=\hsize]{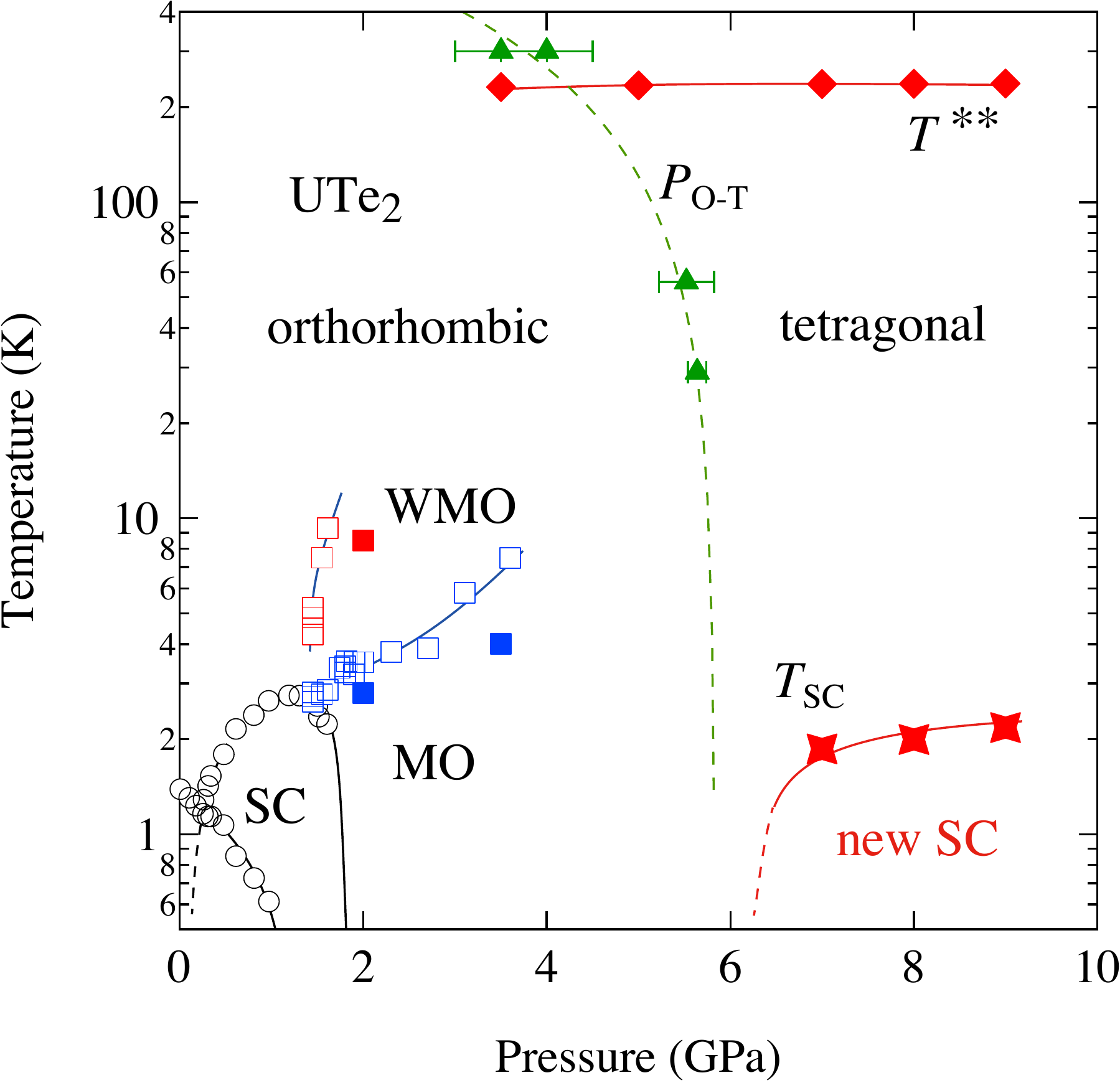}
\end{center} 
\caption{(Color online)
Pressure$-$temperature phase diagram of \UTT\ up to 10 GPa.
Solid lines are guides to the eye.
Solid triangles correspond to the structural phase transition from the orthorhombic to tetragonal structure \Pot\ detected at different temperatures, which were determined upon compression.
The data with closed symbols are determined in this study.
The data points for the low-pressure region (open symbols) are taken from Refs.~\citen{Daniel2019}, \citen{Aoki2020JPSJa}, and \citen{AokiJPSJ2021}.
The dotted green line is the linear extrapolation of $P_{\rm O\mbox{-}T}$.}
\label{gPHD}
\end{figure}

Figure~\ref{gPHD} shows the pressure$-$temperature phase diagram of UTe$_2$. 
The structural transition from orthorhombic to tetragonal occurs at $P_{\rm O\mbox{-}T}\sim 3.5-4\,{\rm GPa}$ at room temperature,
and $P_{\rm O\mbox{-}T}$ shifts to the higher pressure, $P_{\rm O\mbox{-}T} \sim 5.5\,{\rm GPa}$ at 29 and 56 K.\cite{Honda_meeting,Honda2022prep}
Here, $P_{\rm O\mbox{-}T}$ was determined from pressure scans in the XRD experiments at constant temperatures with a small step of pressure increase (approx. 0.5 GPa).
%(ex. $\le$ 1 GPa).
The error bars for the pressure are based on the pressure before and after the XRD measurement.
At low pressures above $1.5\,{\rm GPa}$, $T_{\rm MO}$ corresponding to a magnetic order increases rapidly with pressure, and it disappears above $3.5\,{\rm GPa}$.
%We recall that there exist multiple superconducting phases under high pressure, where the pairing symmetry is not clear yet. Moreover, pressure-induced anomaly, most probably, related to an antiferromagnetic transition, has been observed above 1.5 GPa. 
New anomalies denoted by $T^{\ast\ast}$ start appearing at and above 3.5 GPa, and are almost unchanged up to $9\,{\rm GPa}$.
Superconductivity reappears at around $2\,{\rm K}$ in the tetragonal phase, where the electronic state is drastically changed without the strong electronic correlations, as demonstrated by resistivity curves and the low $H_{\rm c2}$.

%%%%%%%%%%%%%%%   fig05  %%%%%%%%%%%%%%%%%%%%%%%%
%%%%%%%%%%%%%%%%%%%%%%%%%%%%%%%%%%%%%%%

%%%%%%%%%%%%%%%   fig06  %%%%%%%%%%%%%%%%%%%%%%%%
\begin{fullfigure}[htb]
\begin{center}
\includegraphics[width=0.6\hsize]{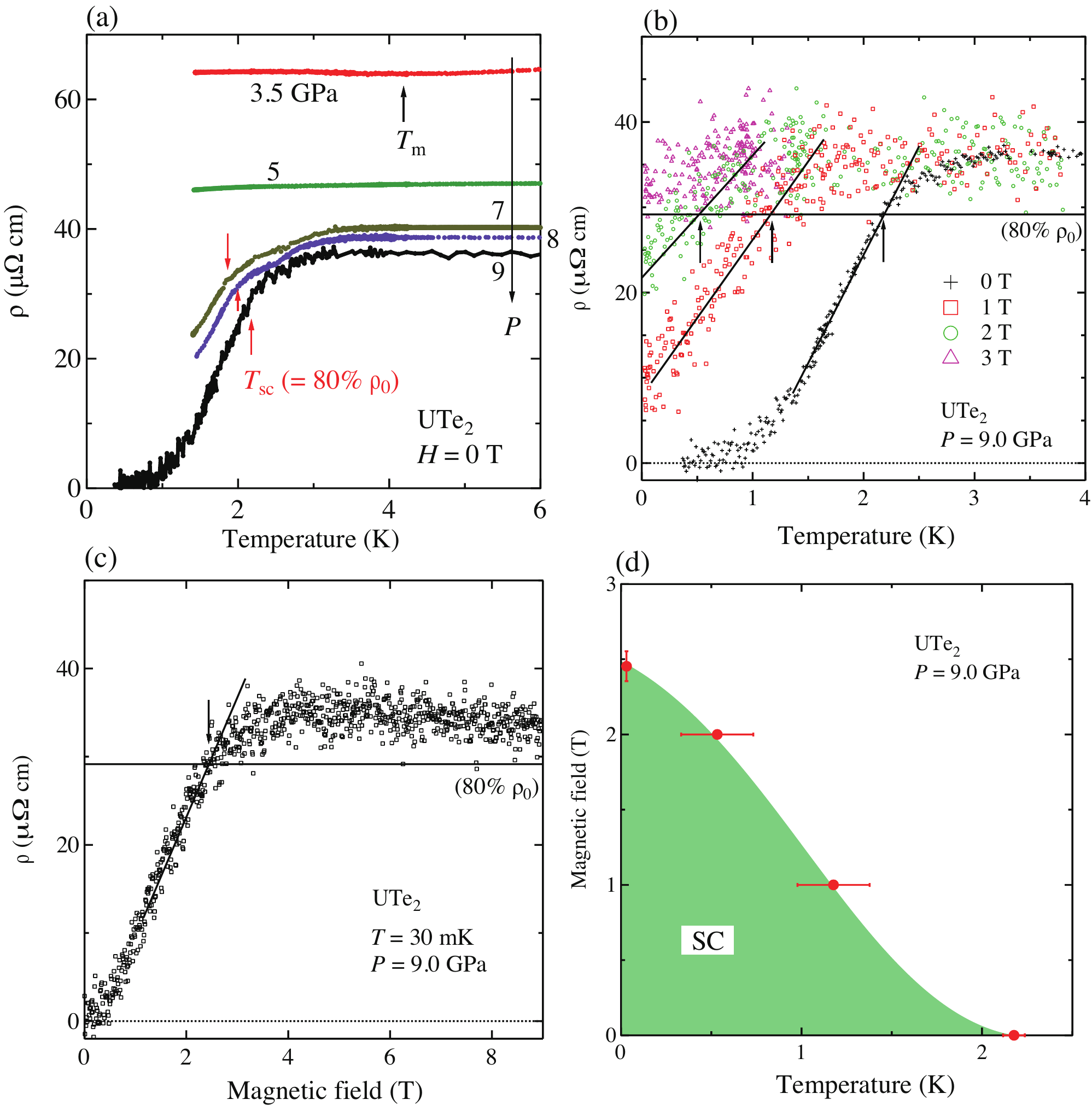}
\end{center} 
\caption{(Color online) 
\DA{(a) Low-temperature resistivities at different pressures above $3.5\,{\rm GPa}$ in UTe$_2$.
(b) Low-temperature resistivities at different fields at $9\,{\rm GPa}$.
(c) Field dependence of electrical resistivity at the lowest temperature, $30\,{\rm mK}$ at $9\,{\rm GPa}$.
(d) Superconducting phase diagram at $9\,{\rm GPa}$. Here, $T_{\rm sc}$ is defined by the temperature and $H_{\rm c2}$ by the magnetic field, at which the resistivity is reduced to $80\,{\%}$ of the normal state resistivity.}
}
\label{HPSC}
\end{fullfigure}
%%%%%%%%%%%%%%%%%%%%%%%%%%%%%%%%%%%%%%%

%%%%%%%%%%%%%%%   fig0X  %%%%%%%%%%%%%%%%%%%%%%%%
%%%%%%%%%%%%%%%%%%%%%%%%%%%%%%%%%%%%%%%

\section{Discussion}

\subsection{Pressure-induced structural phase transition}

We discuss the pressure-induced structural transition at around $P_{\rm O\mbox{-}T}\sim 3.5-4\,{\rm GPa}$. 
The observed diffraction patterns above $P_{\rm O\mbox{-}T}$ are quite different from those in the orthorhombic phase. 
The diffraction pattern is found to be reproduced with two domains of the body-centered tetragonal (BCT) lattice with $a \sim 3.98\,{\rm \AA}$ and $c \sim 9.80\,{\rm \AA}$,
taking into account the extinction rule.

\begin{figure}[htb]
\begin{center}
\includegraphics[width=\hsize]{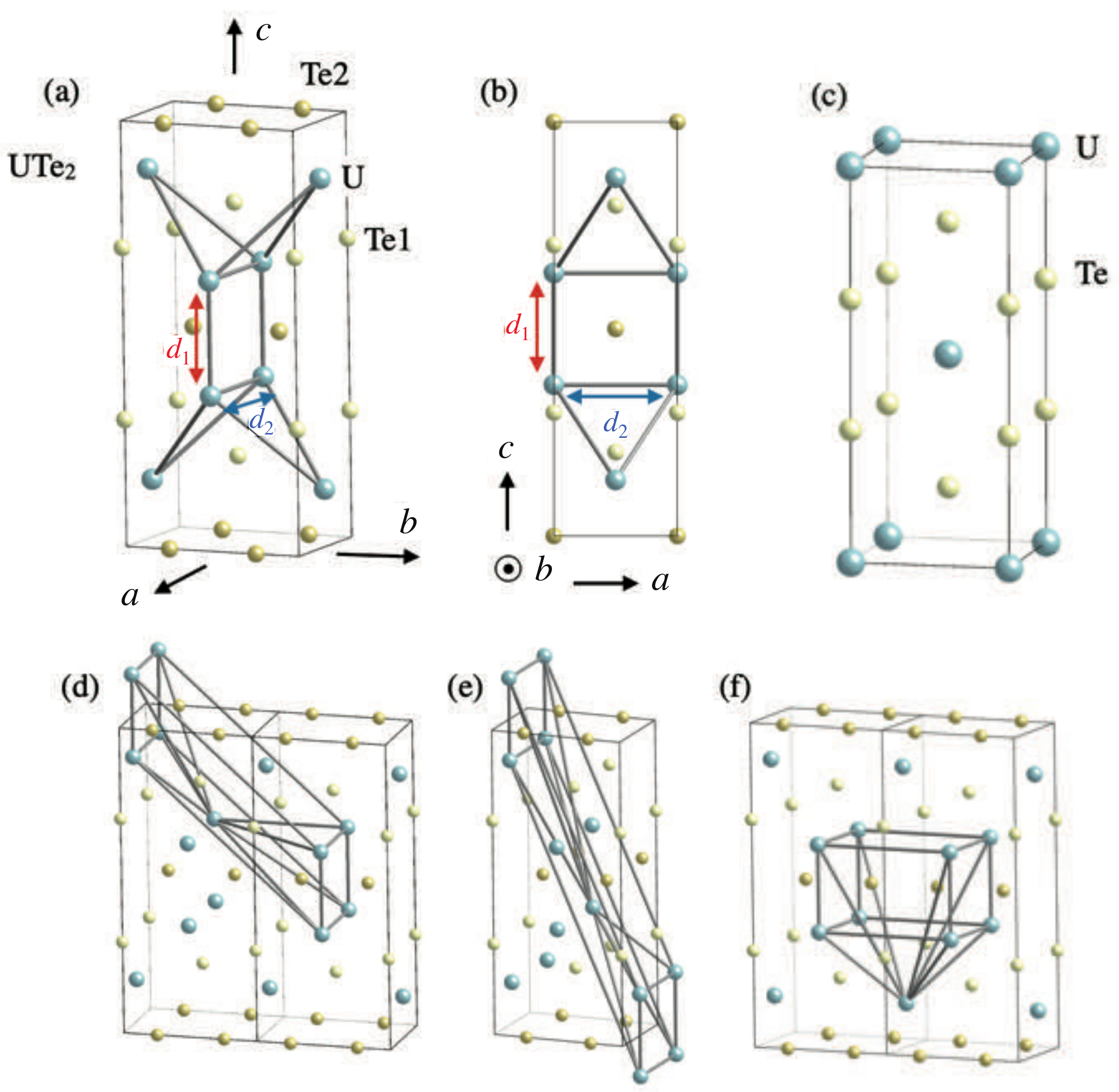}
\end{center} 
\caption{(Color online) 
(a) Orthorhombic structure with the space group $Immm$ at low pressure.
(b) A view from the $b$-axis. The distances of the 1st and 2nd nearest neighbors of U atoms are $d_1=3.78\,{\rm \AA}$ and $d_2=4.17\,{\rm \AA}$, respectively.
(c) Tetragonal structure with the space group $I4/mmm$ at high pressure.
(d)$-$(f) Three possible unit cells for the transformation into the body-centered tetragonal structure are displayed by bonds between U atoms.}
%Schematic drawing of (a) the crystal structure of low-pressure orthorhombic structure, (b) transformation from $Immm$ to $I4/mmm$, and (c) the high-pressure candidate tetragonal structure ($I4/mmm$), where red balls and white balls indicate uranium and tellurium atoms, respectively.
\label{CSchange}
\end{figure}

The volume of the conventional unit cell at ambient pressure is 357 ${\rm \AA}^{3}$ with $Z$ = 4, whereas, above $P_{\rm O\mbox{-}T}$, that of the tetragonal lattice is about $149\,{\rm \AA}$. This corresponds to $Z \sim2$ realized in the high-pressure tetragonal structure and is consistent with the BCT lattice.
Thus, the possible space groups of the high-pressure structure are $I4/mmm$ or its subgroups. Here, we simply take $I4/mmm$ as it is the symmetry of the lattice. It is natural to assume that U atoms are located at the lattice points, namely, at a site 2$a$ (0 0 0).
Given that the Te atoms occupy a crystallographically equivalent site, the position would be limited to site 4$c$ (0 $\frac{1}{2}$ 0), 4$d$ ($\frac{1}{2}$ 0 $\frac{1}{4}$), or 4$e$ (0 0 $z$). Among them, the U$-$Te distance is 1.99 $\AA$ for Te at the 4$c$ site and the Te$-$Te distance is 2.81 $\AA$ for Te at the 4$d$ site, which seems very small considering the ionic radius and strong Coulombic interaction, respectively. 
On the other hand, the Te in the 4$e$ site, with $z ~\sim$ $\frac{1}{3}$, gives the U$-$Te and Te$-$Te distances of about 3.2 $\AA$, which appears to be more reasonable than in the other cases above. The simulated XRD pattern assuming this structure reproduces well the powder XRD pattern of the high-pressure phase, as shown in Fig.~\ref{powXRD}(b). The crystal structures of the low-pressure (LP) orthorhombic and high-pressure (HP) tetragonal phases are depicted in Figs.~\ref{CSchange}(a) and \ref{CSchange}(c), respectively.
Note that the structural phase transition from the ambient-pressure $Immm$ phase of UTe$_{2}$ into a ten fold coordinated, where Te is at the 4$e$ site, high-pressure $I4/mmm$ phase at 9 GPa is suggested by theoretical calculations.\cite{Hu2022}

For these drastic changes in the crystal structure and lattice parameters, large atomic displacements are inevitably required. 
Three possible unit cells that could be transformed into the body-centered tetragonal structure are shown in Figs.~\ref{CSchange}(d)$-$\ref{CSchange}(f).
In all cases, the $ab$-plane for the tetragonal structure is based on the close distances for the 1st and 2nd nearest neighbors, $d_1\sim 3.78\,{\rm \AA}$ and $d_2\sim 4.17\,{\rm \AA}$, respectively.
In Fig.~\ref{CSchange}(d), the rung denoted by $d_1$ should be tilted. Moreover, the atomic position must move to the $a$-axis to obtain the tetragonal structure.
The direction for the new $c$-axis in the tetragonal structure approximately corresponds to the reciprocal $[011]$ direction in the orthorhombic structure.
Note that the reciprocal $[011]$ direction ($24\, {\rm deg}$ tilted from the $b$- to $c$-axis) is different from the $[011]$ direction in real space ($24\,{\rm deg}$ tilted from the $c$- to $b$-axis).
In Fig.~\ref{CSchange}(e), the rung should also be tilted, but no atomic displacement is required along the $a$-axis.
The new $c$-axis approximately corresponds to the $[011]$ direction in real space.
In Fig.~\ref{CSchange}(f), a U atom largely moves along the orthorhombic $c$-axis to obtain the body-centered tetragonal structure.
This is a rather simple displacement of U atoms,
indicating that the orthorhombic $b$-axis switches to the tetragonal $c$-axis.

Although we cannot determine which atomic displacement is correct,
we speculate that the [011] or reciprocal [011] directions are important and they are related to the structural instabilities.
Compared with the crystal structures in LP and HP phases focusing on the U position, the rectangular arrangement of U atoms in the $ac$-plane in the LP phase seems to correspond to the basal plane in the HP tetragonal phase. 
In this structure transformation, the crystal along the $a$-axis maintains its direction and the $ac$-plane including the U rectangular in the $ac$-plane tilts about 25 deg from the $b$ to $c$-axis.
With this transformation, \DA{the $[011]$ direction in the reciprocal space} in the LP orthorhombic phase most likely turns to the [001] axis in the HP tetragonal axis. 
It means that \DA{the reciprocal [011] direction} in the LP phase has been already potentially important in the stability of the crystal structure. 
Note that the above-mentioned transformation is consistent with the change of the diffraction patterns between Fig.~\ref{Laue}(b) and \ref{Laue}(c). 

Since our experiments were carried out on single crystals, not all the diffraction peaks from the tetragonal structure can be obtained, making it difficult to determine the atomic position of Te precisely. 
Several prototype compounds crystallize in the tetragonal $I4/mmm$ type, with the AB$_2$ composition. 
For example, MoSi$_2$, WSi$_2$, and WGe$_2$ are typical compounds, where the $c/a$ ratio is quite similar to that in the pressure-induced HP phase in \UTT, and $z$ parameters are 0.335 in these cases.~\cite{Nowotny1952,Crist1993,Agoshkov1981}
From the theoretical calculation, Hu et al. predicted $z$ = 0.337.\cite{Hu2022}
Here, we assume the $z$ parameter to be 0.335 in the HP phase in UTe$_2$ based on the $z$ parameter of the prototype compounds, which is close to the theoretical prediction.
%Here, we assume the $z$ parameter to be 0.335 in the HP phase in \UTT\ as for the proto-type compounds and plotted the pressure dependence of the nearest neighbor U distance and estimated several U-Te distances in Fig.~\ref{ATdist}(a) and ~\ref{ATdist}(b), respectively.
The pressure dependence of the 1st and 2nd nearest neighbor U distances is depicted Fig.~\ref{ATdist}.
%Note that the next nearest neighbor distance of U atoms in the low-pressure orthorhombic phase corresponds to the lattice parameter of the $a$-axis shown in Figs.~\ref{abcDEP}(a) and ~\ref{AVDEP}(a). 
%The distances between U and the ligand Te atoms become longer and unified. In the low-pressure orthorhombic phase, the nearest Te-Te distance is about 3 \AA\ which is in between the Te(2) atoms along the $b$-axis, where the strongest Coulomb repulsion is expected. Indeed, the $b$-axis in LP phase is the hardest axis. It seems that this structural phase transition is significantly mediated by the Coulombic interactions in the geometric arrangement between Te ions. 
The diverse inter-atomic distances become simpler in the high-symmetry structure of the $I4/mmm$ tetragonal phase. Note that UTeAs, UTeP, and UTeGe form the tetragonal structure with the space group $I4/mmm$. However, the $c/a$ ratio is much larger, and four molecules exist in the unit cell ($Z=4$).

It is worth mentioning that the $c$-axis in the high-pressure phase corresponds to the reciprocal [011] direction in the original low-pressure structure, which is tilted by 24 deg from the $b$- to $c$-axis. 
The field-reentrant superconductivity is observed around this field direction above the metamagnetic field $H_{\rm m} \sim 40\mbox{--}50\,{\rm T}$. 
According to the band calculation of GGG+$U$ ($U=2\,{\rm eV}$) based on the 5$f$-itinerant model, the maximal and minimal cross-sectional areas of the electron Fermi surface coincide with the field direction to the reciprocal [011] direction, which is the so-called Yamaji angle. 
The increase in the density of states is generally expected under a magnetic field at the Yamaji angle, and indeed, in \UTT, the specific heat increases with the field, showing a positive jump at $H_{\rm m}$.~\cite{Miyake2021}
Furthermore, the drastic change in the Hall resistivity is also found for the field direction along the $b$-axis and the reciprocal [011] direction.~\cite{Niu2020,Helm2022}

This suggests that Fermi surface instabilities play a role in field-reentrant superconductivity around the $[011]$ direction above $H_{\rm m}$. 
The principal axis switching to the original reciprocal [011] direction can be another indicator of these Fermi surface instabilities.
One can thus expect further singularities along the [011] or reciprocal [011] direction attributable to uniaxial stress, thermal expansion, and ultrasound experiments.

It is interesting that the flat crystal surface perpendicular to the reciprocal $[011]$ direction, as well as the $[001]$ direction, appears often on as-grown single crystals obtained by the chemical transport method and by the flux method.
The natural crystal surface is generally related to the growth speed, which could be affected by the crystal structure.
This may indicate a precursor for the singularity of the reciprocal [011] direction.

\subsection{Pressure-induced electronic phase transitions}

Two pressure-induced phase transitions appear in \UTT\ for its electronic states in the high-pressure tetragonal phase. One is characterized by the kink of the electrical resistivity at $T^{\ast\ast}=230\,{\rm K}$ and the other one is a superconducting phase transition at around 2 K. 
The former transition was also observed in a previous study by Vali\v{s}ka et al.~\cite{Valiska2021PRB} 
In their study, they observed the $T^{\ast\ast}$ anomaly for the geometry $H \parallel b$, which causes internal strain along the $b$-axis ($u \parallel b$), with their semi-hydrostatic Bridgman anvil cell, but no anomaly was observed for the geometry $H \parallel c$ ($u \parallel c$), suggesting that the $T^{\ast\ast}$ anomaly is sensitive to the uniaxial stress. 
The hydrostaticity of the pressure in our Palm Cubic anvil-type pressure cell is guaranteed by the design itself independent of the pressure-transmitting medium. Therefore, the $T^{\ast\ast}$ anomaly can be favorable to the compression along the $b$-axis in the LP phase. 
When the unit cell is compressed along the $b$-axis, the atomic displacement occurs to avoid strong Coulomb repulsion, particularly between the Te(2) sites mentioned above. 
%The mechanism of the transition at $T^{\ast\ast}$ is unclear so far. 
%But we speculate the transition might be the magnetic one. 
Several uranium pnictides and chalcogenides exhibit high magnetic ordering temperatures.~\cite{Sechovsky1998}
Thus, it is not surprising that the magnetic order could occur at high temperatures in UTe$_2$. 
The nearest neighbor uranium distance in the high-pressure phase of \UTT\ reaches 3.9 \AA, which is sufficiently larger than the so-called Hill limit.
Therefore, it could be possible that the $T^{\ast\ast}$ anomaly is attributed to a switch from the 5$f$ single-site behavior to the coupled 5$f$ array.
Determining the nature of the transition is the primary task for future study. The resistivity anomaly resembles a conventional 2nd-order phase transition in a ferromagnet (or antiferromagnet without a superzone boundary effect - possible for the suggested structure assuming an AF coupling within one unit cell with the magnetic propagation vector $k$ = (0 0 0)). However, without further study, we cannot exclude other options such as a valence change at the pressure-driven transition.
%it is a high probability that the \TSTA\ anomaly is attributed to the magnetic ordering. Then, another question appears: Why does the new pressure-induced superconductivity appear below 2 K in such a robust magnetic state? 

\begin{figure}[htb]
\begin{center}
\includegraphics[width=0.8\hsize]{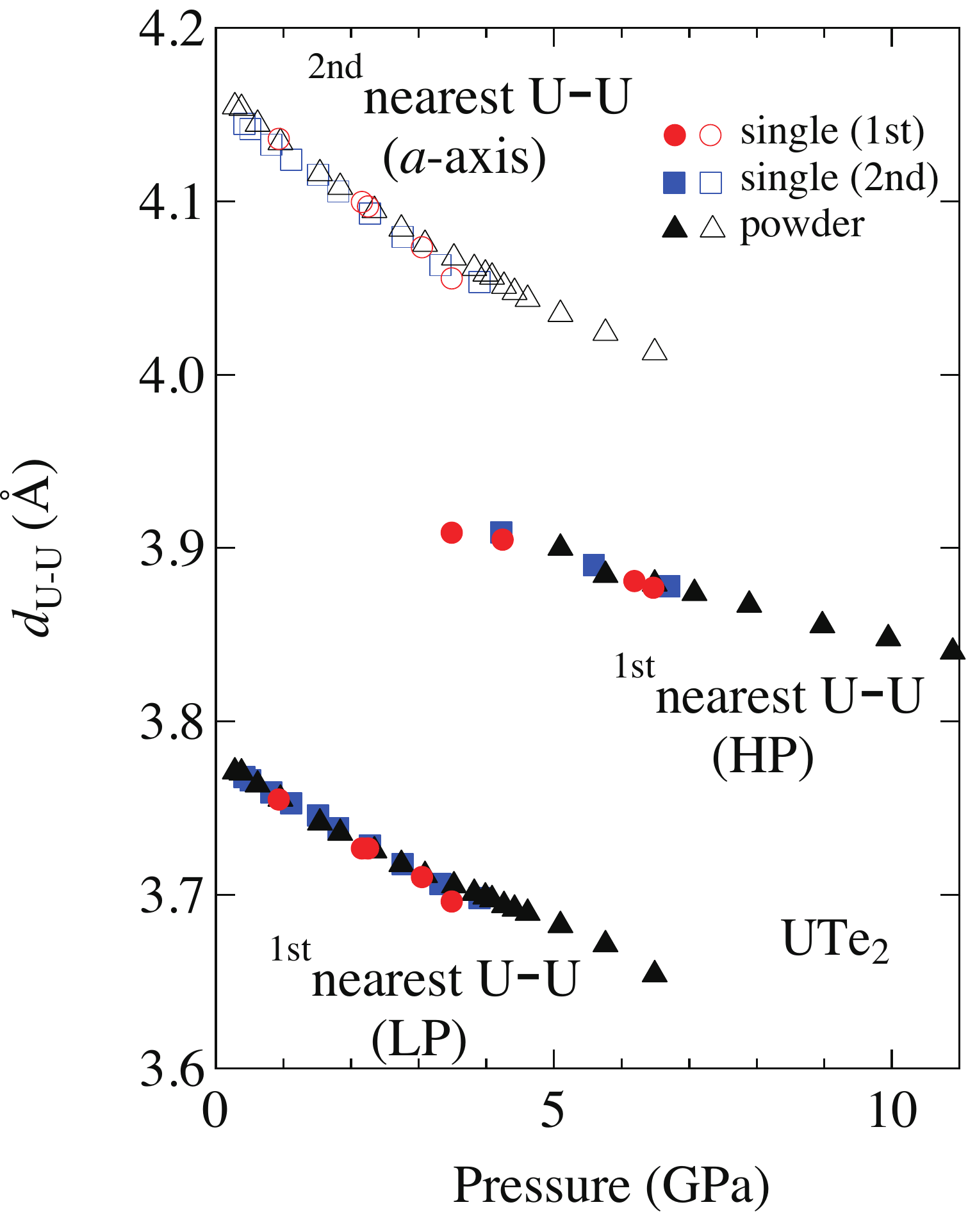}
\end{center} 
\caption{(Color online) 
\DA{Pressure dependence of the distance of the 1st nearest neighbor of U atoms, $d_1$ (closed symbols). That of the 2nd nearest neighbor, $d_2$, in the low-pressure phase is also shown (open symbols). The results of single crystals, (1st run and 2nd run) and powder samples are shown by circles, squares, and triangles, respectively.}
}
\label{ATdist}
\end{figure}

It is clear that the observed superconductivity at high pressures cannot be attributed to the heavy electronic state, as shown by the low \Hc (0) and small $A$ coefficient. 
The drastic change in electronic states associated with the structural change may result in a new superconducting state without 5$f$-electron correlation.
A key point is that the distance between the nearest neighbors of U atoms suddenly increases at the critical pressure from orthorhombic to tetragonal structure as shown in Fig.~\ref{ATdist}, despite the fact that volume decreases with pressure;
this leads to the non trivial results of 5$f$ occupancy, which jumps to $4\,{\%}$ higher than that at ambient pressure.

Similar conclusions have recently been given in reports on pressure studies~\cite{Huston2022}, where the transition from the orthorhombic to tetragonal is reported at around $5\,{\rm GPa}$ under the quasi-hydrostatic condition, associated with an increase in the bulk modulus by $45\,{\%}$, which reaches a value above the critical pressure of the structural transition.
This is close to  a full valence state and the spectacular boost of $d_{\rm U\mbox{-}U}$ by $8\,{\%}$.
The value of $d_{\rm U\mbox{-}U}$ is even larger than that at ambient pressure. 
Furthermore, a new key result is that a trivalent state of the U atom recovers at high pressures in the tetragonal structure.~\cite{Wilhelm2022}

Qualitatively, these observations are in agreement with theoretical calculations~\cite{Hu2022}, in which a structural transition from the orthorhombic to tetragonal is predicted at 9 GPa; 
the main pressure effect is to boost the chemical bonding of a pair of Te electrons. 
A large bulk modulus is also predicted just above the critical pressure. 
Quantitative differences in the critical parameters, ex. $P_{\rm O\mbox{-}T}$, between experiments and theory, are most likely the cause of the difficulty in the treatment for the strong correlation effect in \UTT. 
%From our resistivity measurements SC in the tetragonal phase has most its low pressure heavy fermion character in the vicinity of Kondo lattice insulating phase , of complex magnetic interactions and valence instability .

\section{Summary}
We have performed XRD experiments using single crystals and powder samples to clarify the pressure effect of the crystal structure as well as the electrical resistivity measurements at pressures up to 9 GPa. 
An X-ray diffraction study under high pressure revealed the anisotropic linear compressibility of the unit cell, up to 3 GPa, and a pressure-induced structural phase transition above around 3.5$-$4 GPa at room temperature. 
The body-centered orthorhombic crystal structure with the space group $Immm$ transformed into a body-centered tetragonal structure with the space group $I4/mmm$. 
In addition, we have observed a drastic change in the electronic states accompanied by the structural phase transition and the reappearance of superconductivity above 7 GPa, which does not correspond to  heavy fermion superconductivity.
This could be linked to a relatively rare case, where the U$^{3+}$ configuration is recovered at high pressures.~\cite{Wilhelm2022}

%%%%%%%%%%%%%%%   fig07  %%%%%%%%%%%%%%%%%%%%%%%%
%%%%%%%%%%%%%%%%%%%%%%%%%%%%%%%%%%%%%%%

%%%%%%%%%%%%%%%   fig08  %%%%%%%%%%%%%%%%%%%%%%%%
%%%%%%%%%%%%%%%%%%%%%%%%%%%%%%%%%%%%%%%

%\begin{acknowledgment}
\section*{Acknowledgement}
%\acknowledgment

The authors would like to thank W. Knafo, L. Havela, F. Wilhelm, and J. P. Sanchez for fruitful discussions, and S. Nagasaki for the technical support.
This work was performed with the approval of the Japan Synchrotron Radiation
Research Institute (JASRI) (Proposal Nos. 2020A0741, 2020A0740, and 2021A1527) and under the Inter-University Cooperative Research Program of the Institute for Materials Research, Research Center for Nuclear Materials Science of Tohoku University (Proposal Nos. 202012-IRKAC-0017, 202012-IRKAC-0056, 202112-IRKAC-0029, and 202112-IRKAC-0041).
This work was financially supported by KAKENHI
(JP19K21840, JP19H00648, JP20K03827, JP20K20889, JP20H00130, JP20KK0061, JP22H04933, JP22K03516).

%a Grant-in-Aid for Challenging Research (Exploratory) (JP19K21840), Scientific Research (A) (JP19H00646), and (B) (JP20K03827) from the Japan Society for the Promotion of Science.

%\end{acknowledgment}

\end{document}